# *To what extent are multiple pendulum systems viable in pseudo-random number generation?*

By Matthew Sigit

# Table of Contents

## 1. Introduction

More so than any other genre, sandbox games give a feeling of infinite possibilities through a combination of their simple rules and ever-changing environment interwoven into a complex web of interactions. One such game that I often enjoy exploring is Minecraft, in which "Seedfinding" - the process of finding interesting generated worlds - has been a passion project of mine since multiple years ago. This hobby has led me to discover the flaws of its generation, which are derived from the programming language Java.

It is through this that I learnt about the usage of random numbers in our day-to-day lives. Random numbers, which have extensive applications in computers, can be categorized into two types. The first includes those that are truly random, taken from the usage of chaotic systems and probabilistic events rather than deterministic, such as time, quantum particles, radiation and other sources of entropy. Some truly random numbers depend on the process's unpredictability, while others will use the inaccuracy that results from taking precise measurements (such as quantum particles due to Heisenberg's uncertainty principle, which states that "the position and the velocity of an object cannot be measured exactly, at the same time, even in theory"[1]); nevertheless, these are inherent to physical processes. This means that these truly random numbers require real world equipment to measure and are thus able to become quite costly. Regardless, some companies may opt to use these random numbers for their own purposes - an example being Cloudflare using images of the floating material in lava lamps[2] for security purposes[3].

---

[1] Encyclopedia Britannica. u*ncertainty principle | Definition & Equation*. [online] Available at: <https://www.britannica.com/science/uncertainty-principle>.

The second category, which this investigation will focus on, are those created through an algorithm and seem to be random, aptly named pseudorandom. Pseudorandom numbers are created by algorithms in computers known as Pseudorandom Number Generators, which I will refer to as PRNGs throughout this essay. PRNGs, as they can be created through computers in milliseconds, are often used in favor of their more expensive yet more "random" counterparts. Their viability is assisted by the fact that they typically require less energy than real world number generators like the aforementioned lava lamps, which must take hundreds of images and transmit them, leading to a persisting strain on bandwidth.

The importance of PRNGs are cemented by the resulting widespread use of PRNGs in fields such as security, games and gambling. In security, random number generators are essential because of the nature of computers and their processing abilities. Random numbers are needed to generate keys, as well as initialization vectors and nonces - numbers used to ensure the security of private communication by stopping potential hackers from replaying data[4].

In games and gambling, random numbers are needed in order to create a sense of unpredictability and make the game seem less deterministic through the introduction of chance-based events and items. Random numbers are also used for generating maps. Without them, games would play out the same every time and, for some, would ruin their replay value.

---

[2] Selden, Annie & Selden, John.The College Mathematics Journal; Washington Vol. 29, Iss. 1,  (Jan 1998): 77.
[3] Cloudflare. *Lava Lamp Encryption | How Do Lava Lamps Help With Internet Encryption?.* [online] Available at: <https://www.cloudflare.com/learning/ssl/lava-lamp-encryption>.
[4] Hypr. *What is a Cryptographic Nonce? | Definition & Equation*. [online] Available at: <hypr.com/nonce>

On the other hand, depending on the algorithm and the seeds - the initial numbers that computers will place into the algorithm - these PRNGs will have varying amounts of success in creating random sequences of numbers.

This topic is worthy for study because, despite the widespread use and clear market, there are issues with pre-existing random number generators. In the short time that PRNGs have existed, there have been prominent failures in security due to a misuse of random numbers. For example, an analysis conducted in 2012 displayed that, from millions of keys, they were able to factorize and gain access to thousands of accounts due to a failure in seeding the initial values, thus creating important keys easily factored by Euclid's algorithm alone[5]. The ramifications of such a large proportion of keys potentially containing sensitive information places into question the security of the information available on the internet as a whole.

Multiple-pendulum systems may serve to be an interesting source of random numbers. This system contains pendulums attached by their ends, forming a chain that is often referred to as one of the most chaotic simple systems[6]. Chaotic, in this sense, refers to a high sensitivity to initial conditions resulting in large differences at later states of a deterministic non-linear system. Multiple pendulum systems, which "exhibit a growth of uncertainties which is exponential"[7] are included in this category and thus may prove to be a more robust method, not only because of its viability of creation in computers (which are inherently deterministic[8] and thus can model the

---

[5] Lenstra, Arjen K. et al. International Association for Cryptologic Research. *Ron was wrong, Whit is right.* Published 2012. [online] Available at: <https://eprint.iacr.org/2012/064.pdf>
[6] Calvão, A.M. & Penna, T.J.P. European Journal of Physics. *The Double Pendulum: a numerical Study.* Published 2015. [online] at https://iopscience.iop.org/article/10.1088/0143-0807/36/4/045018/meta
[7] American Journal of Physics 60, 491 (1992); https://doi.org/10.1119/1.16860
[8] Rubin, Jason M. MIT School of Engineering. *Can a computer generate a truly random number?.* Published 2011. [Online] at

system) but also as small changes in the seed potentially will not have the same issues as the above keys. Despite this, the usage of pendulum systems in cryptography is not frequently discussed.

This investigation will explore the generation of random numbers in Java and their flaws. It will also contain the results of creating and optimising my own PRNG using multiple-pendulum systems and comparing mine against existing library classes, in order to answer the question:

**To what extent are multiple pendulum systems viable in pseudo-random number generation?**

## 2. Background Research

### 2.1 Existing deficiencies in Java

As aforementioned, there is an issue with java's random number generator - it is linear. This generator, known as a "Linear Congruential Generator (LCG)", is one that derives a pseudo random number from an equation ax+b, where a and b are constants, and x is the seed. Although this may lead to less memory and time usage, for similar values of the seed, the end result will be very similar as well, with the only difference being the difference in seeds multiplied by constant A, which is readily available for anyone to read on the source code[9]. As a result, people are able to derive the original value of seeds for any randomly generated item, and thus manipulate them to their advantage - for example, finding seeds with infinitely repeating terrain due to using seeds which are perfect factors of these numbers, leading to an unchanging value in modular arithmetic. Furthermore, the similar initial seeds for two different blocks, with one

---

https://engineering.mit.edu/engage/ask-an-engineer/can-a-computer-generate-a-truly-random-number

[9] Katleman. Open JDK. *Random.Java* (Source code). Published 2014. [Online] at: https://hg.openjdk.java.net/jdk8/jdk8/jdk/file/tip/src/share/classes/java/util/Random.java

being much more difficult to obtain/observe than the other, may lead to a method that connects the appearance of one block to the other (i.e. if the more common block is generated, players are able to use the similar initial values to find the rarer block)[10].

To this end, Java has made a SecureRandom class which produces cryptographically secure random numbers, which is mentioned by the Random class to be a "cryptographically secure pseudo-random number generator for use by security-sensitive applications"). This class uses a SHA1-PRNG[11] - standing for a Secure Hash Algorithm Pseudo-Random Number Generator, which was frequently used in the past for security purposes.

The java random function and SecureRandom function will set a useful baseline for this test, as they are widely used - being the default random number generator in one of the largest languages with applications in cryptography and games. Furthermore, it is written in the same language as my program, and thus will provide a better metric as the same language implies similar resource usage.

## 2.2 Multiple Pendulum Systems

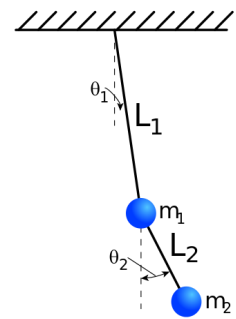

The importance of multiple pendulum systems in teaching students about chaos cannot be understated; models are often used in education due to its chaos despite its simplicity[12]. A crucial fact about these systems that will be

---

integral for the experiment will be that its angles cannot b e predicted[13] through analytical integration of its lagrangian equations, which will assist in security as its seed cannot be easily reversed through its end results.

This fact is incredibly important, as the acceleration of objects is based on the trigonometric functions of their angles. As the only energy sources present in the pendulum system are the gravitational potential and the kinetic energy of the pendulums, and the gravitational force on pendulums in terms of their component forces (the vertical being the cosine of the angle multiplied by the magnitude, and the horizontal being the sine[14]), the angles in calculation are important.

The motion of double pendulums can be traced through a set of ordinary differential equations, which are as follows:

$$\theta_1'' = \frac{-g(2m_1 + m_2)\sin\theta_1 - m_2 g \sin(\theta_1 - 2\theta_2) - 2\sin(\theta_1 - \theta_2) m_2 (\theta_2'^2 L_2 + \theta_1'^2 L_1 \cos(\theta_1 - \theta_2))}{L_1 (2m_1 + m_2 - m_2 \cos(2\theta_1 - 2\theta_2))}$$

$$\theta_2'' = \frac{2\sin(\theta_1 - \theta_2)(\theta_1'^2 L_1(m_1+m_2) + g(m_1+m_2)\cos\theta_1 + \theta_2'^2 L_2 m_2 \cos(\theta_1 - \theta_2))}{L_2(2m_1 + m_2 - m_2\cos(2\theta_1 - 2\theta_2))}[15]$$

\*Comment: There was an issue with creating equations, as they would not fit, and subscripts were not accepted.

This equation is deduced through the Newtonian method, using algebra and Newton's second law applied to the masses. The position of the first pendulum can be derived through the position of the normal and trigonometric equations. The angle between a line parallel to the direction of gravity and the pendulum, and the length of the pendulum leads to an equation for the coordinates, as the difference in x coordinates of the normal and the first pendulum is equal to $L\sin\theta_1$ (i.e. the angle between the downwards normal and the length of the pendulum). The difference in y coordinates can be calculated through $L\cos\theta_1$ as well. Similarly, the position of the second pendulum can be calculated with its angle and length, with the first pendulum acting in place of the fixed end of the first pendulum, where the angle is located.

Because velocity is the derivative of the difference in position, and acceleration is the derivative of velocity with respect to time, then equations for velocity and acceleration can be created through the position equation detailed above. Thus, applying newton's second law

---

[15] My Physics Lab. *Double Pendulum.* [Online] at https://www.myphysicslab.com/pendulum/double-pendulum-en.html

F = ma creates a secondary equation for acceleration, used to create the above equation.

## 2.3 Quantifying Random: The NIST Statistical Test Suite

To measure the efficacy of this type of PRNG, I will be using two results - the randomness, and the resources used to create each random number.

The random number generator's ability to create random numbers will be measured through the NIST statistical test suite. This set of tests, developed after the revolutionary encryption method AES replaced DES[16], measures randomness in sequences of bits (ones and zeros) through factors such as the frequency of repetitions, variations and patterns in the numbers created - the full list is written below.

Tests: Frequency, block-frequency, cumulative sums, runs, longest run, Rank, DFT, non-overlapping, overlapping, universal, approximate entropy, random excursions, random excursions variant, serial, linear complexity.

The measurements of frequency, block frequency, runs etc. are very important for random number generators. If some numbers appear far more than others, this may lead to an error in which some the result is biased for some outcomes - this can have very large implications as small chances may be far larger than anticipated, possibly leading to losses in gambling should a casino attempt to use a frequently-occuring number chain for a high-value outcome.

The test will search through the sequence of numbers fed into it to determine whether a sequence of numbers will pass its tests or not. This can possibly be repeated for potentially thousands of sequences/files to create an aggregate score, which I will then compare to the values received from other PRNGs.

---

[16] Zaman, J K M Sadique Uz & Ghosh, Ranjan. Institute of Radio Physics and Electronics, University of Calcutta. *A Review Study of NIST Statistical Test Suite: Development of an indigenous Computer Package.* [Online] at https://arxiv.org/pdf/1208.5740.pdf

## 2.4 Quantifying Resources

The resources used are easier to measure than the "randomness" of the bit sequences. , as the memory usage and time taken is readily available in program frameworks and through modules inside of the program. The time can be measured using the time module, which can be done by taking the time before and after creating a random number. The memory can be measured through the Runtime module.

# 3. Methodology

## 3.1 Initial Test

The program will be created in Java, and will use the aforementioned equations to create the equations for the pendulum's motion. To simplify the program, the pendulum will be limited to a two-dimensional plane using cartesian coordinates. To create the initial parameters, random numbers from PRNGs will be used to begin with, as well as time.

After these variables and the control variables are set, the program will run a loop to simulate the double pendulum. The number of loops can range for a random number between 1,000 and 10,000, to allow ample iterations for similar initial seeds to diverge in their positions. The final random number will then be taken and converted into binary so that it can be passed into the NIST statistical test suite. To keep the number within the memory size outlined by java, the value will be placed into a file.

Finally, this will repeat until the file has approximately 1 million digits inside of it. It will then be uploaded to a website which hosts the NIST statistical test suite. The program will run to create 10 files.

To calculate the memory, the program will create a (approximately) billion digit number, and subtract the memory usage at the start of the program from the end. Similarly, the program will create a billion digit number and calculate the time taken, then divide to find the average time for a million digits.

For another test, damping will also be implemented, which will reduce the overall energy of the system over time. This will be done to compare the use of random numbers with damping against without. The damping coefficient will be set to 0.9999 - thus, the program will lose 0.01% of its energy for every loop.

## 3.2 Control Variables

The control variables involved will include gravity, set to 9.81, to mimic the Earth's gravity and the relative lengths of strings (set equal). The range for the random values (1000-10000 for loops which represent time and 1-300 for the masses) will also stay the same throughout the experiment. There are other variables in the NIST statistical test suite - these will be set to the base values recommended by NIST[17]. The following values are written/calculated below:

| Test # | Test name | $n$ | $m$ or $M$ | # sub-tests |
|---|---|---|---|---|
| 1. | Frequency | $n \geq 100$ | - | 1 |
| 2. | Frequency within a Block | $n \geq 100$ | $20 \leq M \leq n/100$ | 1 |
| 3. | Runs | $n \geq 100$ | - | 1 |
| 4. | Longest run of ones | $n \geq 128$ | | 1 |
| 5. | Rank | $n > 38\,912$ | - | 1 |
| 6. | Spectral | $n \geq 1000$ | - | 1 |
| 7. | Non-overlapping T. M. | $n \geq 8m - 8$ | $2 \leq m \leq 21$ | 148* |
| 8. | Overlapping T.M. | $n \geq 10^6$ | | 1 |
| 9. | Maurer's Universal | $n > 387\,840$ | | 1 |
| 10. | Linear complexity | $n \geq 10^6$ | $500 \leq M \leq 5000$ | 1 |
| 11. | Serial | | $2 < m < \lfloor \log_2 n \rfloor - 2$ | 2 |
| 12. | Approximate Entropy | | $m < \lfloor \log_2 n \rfloor - 5$ | 1 |
| 13. | Cumulative sums | $n \geq 100$ | | 2 |
| 14. | Random Excursions | $n \geq 10^6$ | | 8 |
| 15. | Random Excursions Variant | $n \geq 10^6$ | | 18 |

[18]

Block Frequency M = 20

Bitstreams = 1

Length = 100,000

Approximate Entropy M < Floor($\log_2$(100000))-5 = 11, m = 10

Serial M < Floor($\log_2$(100000))-2 = 14, m=13

For the sake of simplicity, four tests which increase the length of time too much are excluded from the experiment results: being the Non-overlapping, overlapping, random excursions, random excursions variant and linear complexity tests. These tests increase time, as random excursions and their variants have a combined 26 sub-tests, drastically increasing the amount of time, and the non-overlapping test contains 148 sub-tests, whereas the other tests typically only contain one sub-test.

---

[17] Sýs, Marek & Riha, Zdenek & Matyas, Vashek. Research Gate. *NIST Statistical Test Suite – result interpretation and optimization.* Published 2015. [Online] at https://www.researchgate.net/publication/288984210_NIST_Statistical_Test_Suite_-_result_interpretation_and_optimization

[18] Rukhin, A. , Soto, J. , Nechvatal, J. , Smid, M. , Barker, E. , Leigh, S. , Levenson, M. , Vangel, M. , Banks, D. , Heckert, N. , Dray, J. and Vo, S. National Institute of Standards and Technology, Gaithersburg, MD. *A Statistical Test Suite for Random and Pseudorandom Number Generators for Cryptographic Applications, Special Publication*. Published 2001. [Online] at https://www.nist.gov/publications/statistical-test-suite-random-and-pseudorandom-number-generators-cryptographic-0

## 3.3 Comparison Tests

In order to compare my program against pre-existing PRNGs, I will be creating 10 files for each of the Java random classes as well. In order to do so, I will call the random int function repeatedly on both types, convert the values into binary and append the values until I reach a sequence of numbers approximately 1 million digits long. I will then upload all the files into the NIST statistical test suite, using the same control variables, and measure the memory and time with the same method.

## 3.4 Further Optimization

Many values in the initial test will be set, however, changing some values may lead to a more efficient and/or more "random" result. The addition of damping is already explored above, however the values for gravity, and string lengths can be varied. This can be done through a loop, which will run the initial test to find the values which score better on the NIST statistical test suite.

# 4. Experiment Results

## 4.1 NIST results of initial test and comparisons

Below is a tabular representation of the test names, next to the number of files that completed the test results and received a P value of 1 (pass).

| Test Name | Java Random | Pendulums (Damping) | Pendulums (No Damping) | JavaSecure |
|---|---|---|---|---|
| Frequency | 2 | 4 | 0 | 0 |
| Block Frequency | 0 | 2 | 0 | 10 |
| Cumulative Sums | 3 | 5 | 0 | 9 |
| Runs | 1 | 2 | 0 | 8 |
| Longest Run of Ones | 0 | 7 | 2 | 7 |
| Rank | 3 | 3 | 0 | 6 |
| Discrete Fourier Transform | 4 | 2 | 0 | 9 |
| Universal Statistical | 2 | 4 | 0 | 8 |
| Approximate Entropy | 0 | 3 | 1 | 7 |

| Serial | 0 | 6 | 0 | 8 |
|---|---|---|---|---|
| Overall | 15 | 38 | 3 | 72 |
| Average Memory Usage (KB) per 1 million denary digits | 912 | 1870 | 1890 | 2309 |
| Time per 1 million denary digits | 25 | 23 | 24 | 21 |

## 4.2 Result of Optimization

The optimization showed insignificant differences between changing the gravity, and negative correlation with increasing the length of the first string in relation to the second.

## 4.3 Applications of Findings

Through these findings, we learn that the inclusion of damping is not beneficial in creating random numbers for the allocated time and conditions provided. However, the frictionless multiple pendulum system is more viable in the Java programming language.

The results of these findings may prove useful for computer scientists and the fields in which pseudo-random number generators are applied. The multiple-pendulum system has proven that some instances are able to pass the NIST statistical tests. Should the number generator be applied in different ways, languages and by different people, this could result in potentially cryptographically-secure systems that are based on the physics of pendulums.

In another vein, the position of this pseudo-random number generator as amid the two java random number generators could serve to help many. This program, or more efficient versions which reduce computational time and resources, could be useful as a compromise for those who would like a more "random" number whilst not compromising too much memory and time. Although this program currently does not have as much use as others in creating the most secure hashes for cybersecurity, it can be used in lighter applications, such as game generation, due to its light memory load and superior performance to the basic java random class.

## 5. Further Research

This pseudo-random number generator was chosen for its simplicity, leading to less time and memory being used to conduct experiments. A system with greater emphasis on randomness may use far more pendulums in the system, or have multiple pendulums attached to certain points. This system was also made in two dimensions, meaning that the viability of a three-dimensional multiple pendulum system is unknown.

Furthermore, the NIST statistical test suite is also not the only test that can be used to judge the random number generators. Further research can be done into other tests, such as the Shannon entropy and the DIEHARDER statistical test suites, to observe whether this system continues to prove viability in different tests.

It should also be noted that the further optimization was not completed in the most robust manner possible. Rather, the system attempted to search for the most efficient value for many variables, one at a time, and within a small range, with the assumption that a high randomness score for one variable would mean that the same variable would affect the randomness score positively for other values of other variables (e.g. value X for gravity, which yielded the best result for the downwards force, would also yield the best results when testing for value Y of string length). Due to time and resource constraints, combining the testing of variables would have been impractical, as the resource usage would change exponentially. Another assumption made is that an integer value would create sufficiently random results that would be able to compare against values outside of this range (e.g. non-integer values), which also includes the boundaries set for the randomly generated initial values. Further research can be undergone to prove or disprove whether these assumptions are correct, and, should these be incorrect, to create different values for optimization of efficiency.

A few tests were left out of the experiment for the sake of simplicity. Further resource can be done to explore the merits of this experiment in terms of overlapping, non-overlapping template matchings, random excursions, random excursion variants and linear complexity.

Finally, this paper has only covered the merits of a two-dimensional multiple pendulum system in a cartesian format on java. Further research can be done to study the effects of a third dimension, more pendulums or a change in language on overall performance.

## 6. Conclusion

This paper analyzes the effectiveness of a different type of pseudo-random number generator - a multiple-pendulum based number generator, chosen for its chaotic nature and simplicity. This is done by analysing the numbers through the NIST statistical test suite. The paper analyzes this pendulum system with different values of damping, weights and lengths.

The results show that the pseudo-random number generator without damping is in the median of the two classes it has compared to, in terms of resource usage and score, which may have a variety of applications. The results also show that the model proved to work better when used without reducing the overall force of the system over time. This is presumed to be because, for higher lengths of time, the overall force will tend towards zero, leading to a higher chance of certain numbers appearing.

This paper may be useful in creating an intermediary stage between the random classes in Java. As it has proven that a pseudo-random number generator based on pendulums are more resource intensive in terms of time and memory than the java random, but less than the securerandom class, it may appeal to people who place more emphasis on randomness - enough such that rather than using the basic class, they would prefer a more secure solution, whilst wanting to use less resources.

Despite the limitations in resources, expertise and the language itself, the multiple-pendulum system has a niche in creating random numbers. This will hopefully encourage more computer scientists to consider switching to the simple yet chaotic deterministic system as a baseline for their pseudo-random number generators. This can benefit certain sectors, who may benefit from a generator produced more efficiently through a different language.